\documentclass[twocolumn,prb]{revtex4}
\usepackage{amsmath}
\usepackage{amssymb}
\usepackage{float}
\usepackage{graphicx}
\usepackage{natbib}
\usepackage{bbold}
\usepackage{color}

\DeclareMathOperator{\tr}{Tr}

\newcommand{\ket}[1]{| #1 \rangle}
\newcommand{\bra}[1]{\langle #1 |}

\newcommand{\be}{\begin{equation}}
\newcommand{\ee}{\end{equation}}
\newcommand{\ba}{\begin{eqnarray}}
\newcommand{\ea}{\end{eqnarray}}
\def\tr{\mbox{tr}}

\begin{document}
\title{Probing charge fluctuator correlations using quantum dot pairs}
\author{V. Purohit}
\author{B. Braunecker}
\author{B. W. Lovett}
\email{vp27@st-andrews.ac.uk}
\email{bwl4@st-andrews.ac.uk}
\email{bhb@st-andrews.ac.uk}
\affiliation{Department of Physics and Astronomy, St Andrews University, St Andrews, United Kingdom}
\date{\today}

\begin{abstract}
\noindent
We study a pair of quantum dot exciton qubits interacting with a number of fluctuating charges that can induce a Stark shift of both exciton transition energies. We do this by solving the optical master equation using a numerical transfer matrix method. We find that the collective influence of the charge environment on the dots can be detected by measuring the correlation between the photons emitted when each dot
 is driven independently. Qubits in a common charge environment display photon bunching, if both dots are driven on resonance or if the driving laser detunings have the same sense for both qubits, and antibunching if the laser detunings have in opposite signs. We also show that it is possible to detect several charges fluctuating at different rates using this technique. Our findings expand the possibility of measuring qubit dynamics in order to investigate the fundamental physics of the environmental noise that causes decoherence.
\end{abstract}
\maketitle
\section{Introduction}\label{introduction}

Quantum dots (QDs) are semiconductor heterostructures exhibiting electronic confinement in all three spatial dimensions. As such, a QD is zero-dimensional, and its eigenstates resemble those of a particle in a box.\cite{piab,piab2} They are `artificial atoms' and many properties typical of a discrete energy level spectra have been observed, for example Rabi oscillations.\cite{rabi_exciton,ducas,rabi_exciton2, ramsay10} Qubits may then be represented through a variety of different kinds of particles in QDs, including electron and hole spin or exciton states.\cite{graphene,rabi_exciton,vincenzo, morton10, atature06, brunner09} Such QD exciton qubits have large transition dipoles and interact strongly with an optical field, and therefore QDs make excellent sources of single photons.\cite{singlephoton,atature,kok10} 

\begin{figure}[H]
\centering
\includegraphics[height=50mm]{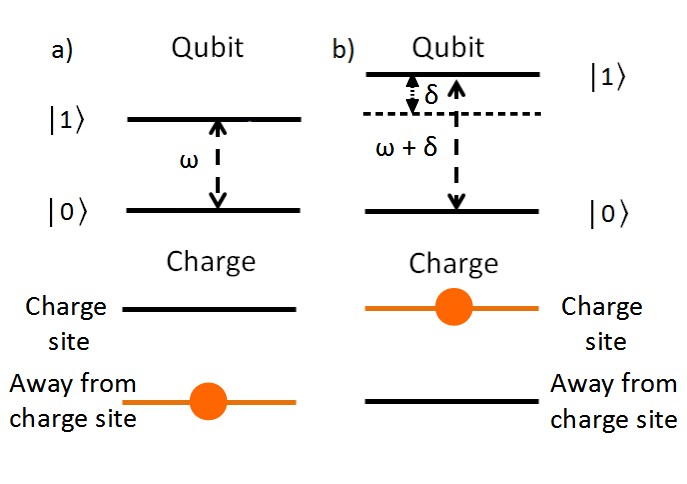}
\caption{The qubit energy shift for an unoccupied (left) and charged (right) trap. The exciton creation energy is denoted by $\omega$ and the charge-qubit interaction strength is $\delta$.}
\label{stark}
\end{figure}

\begin{figure}[H]
\centering
\includegraphics[height=40mm]{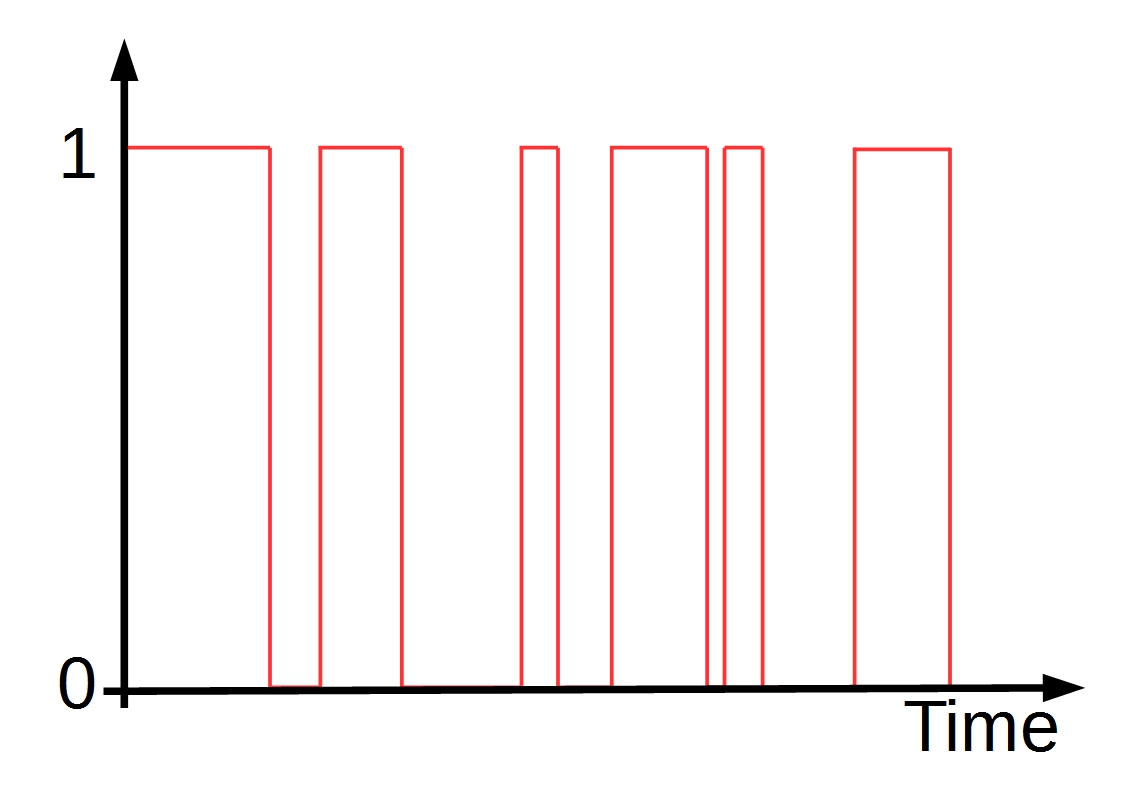}
\caption{Demonstrative graph of telegraph noise. The fluctuating charge switches stochastically between states 1 (charged) and 0 (uncharged).}\label{telegraph}
\end{figure}

A key challenge in assessing the feasibility of any quantum computer realisation is to develop an understanding of decoherence in the system. Though a great deal of work has been done on how individual qubits suffer decoherence, less is known about how correlated noise across multiple qubits arises from different kinds of environment. Any long-range interaction with an environmental disturbance will cause correlated noise channels for relatively closely spaced qubits. One such example is that of fluctuating charges in the vicinity of QDs.\cite{charge1,joynt,charge2,wold,charge4} Semiconductors, by their nature, have a Fermi energy in the band gap, which can be small enough in doped samples that the conduction band can be thermally occupied when the temperature is relatively low. Vacancies or impurities in the crystal structure of a semiconductor lead to local alterations to the band structure, and charges can become trapped in lower lying states. Depending on temperature, such charge traps will randomly switch between empty and full, and they may then modelled as two state fluctuators. In the vicinity of QD excitons, such fluctuators lead to random telegraph noise in the exciton energy, and so in the emitted photon frequency, through the DC Stark shift;\cite{joynt,charge4} see Figs.~\ref{stark} and~\ref{telegraph}.

In this paper, we show that the common, correlated, noise that is generated by charges fluctuating in the vicinity of two QDs can be detected by driving the QDs optically and analysing the subsequently emitted photons. Specifically, we determine the cross correlation function $g^{(2)}(t, t+\tau)$ of the emitted light, and we will show that it can reveal a wealth of information about the nature of the charge environment, including how common it is to both qubits. In some cases we will find that it is possible to determine how many charges interact with the QD states, and at what rate they fluctuate.

In the following section, we present the model of the QDs coupled to both a photonic bath and a common environment of charge fluctuators, and describe the Markovian master equation for the QD-radiation coupling. We also present the transfer matrix method for solving the time evolution of the density matrix. In section ~\ref{methods} we introduce the photon correlation function for two driven QDs. In section~\ref{results}, we present out predictions before summarising in section~\ref{summary}.

\section{Model}\label{hamiltonian}

\begin{figure}[H]
\includegraphics[height=60mm]{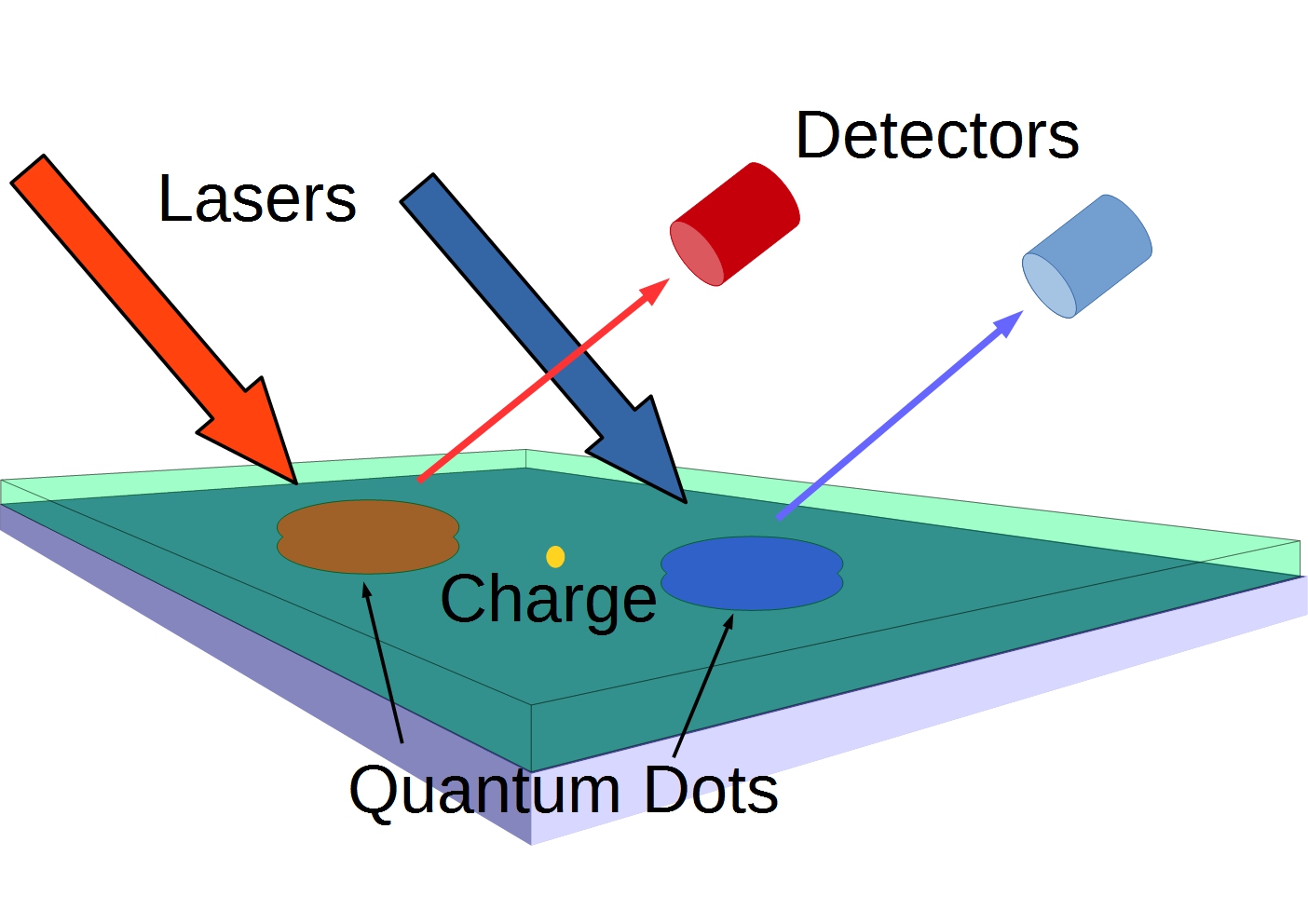}
\caption{Diagram of the experimental set up. The two QDs are driven by lasers of different frequencies. The emitted photons are captured by the detectors, where the Hanbury Brown-Twiss style experiment is performed.}\label{schem}
\end{figure}

We consider two driven QDs, that are modelled as two level systems with different energy spacings. These emit photons of different frequencies, which can be measured by two time-resolving detectors. The QDs are driven by lasers with different frequencies to match their respective resonances, though each can be slightly detuned from this condition. In addition, these uncoupled QDs both interact with a common environment that takes the form of a limited number of charge fluctuators that will also be represented by two level systems; the situation is shown  in Fig~\ref{schem}. The fluctuators will be treated as classical objects with no coherence between the charged and uncharged states. The charges affect the qubits via the Coulomb interaction, inducing a DC Stark shift. We will neglect any dielectric screening effects, which would anyway simply introduce a renormalisation of the effective distance between qubits and fluctuators. 

The Hamiltonian is written as (for full details of the Hamiltonian construction see Appendices~\ref{appa} and \ref{appb}):
 \ba
 \label{eq1}
 H &=& \sum_{i}\frac{\omega_{i}}{2} \sigma_{z,i} + \sum_{k}\theta_{\bf k}a_{\bf k}^{\dagger}a_{\bf k} +\sum_{j}\frac{\xi_{j}}{2}\eta_{j}\nonumber\\
  &&+ \sum_{{\bf k},i}\zeta_{i, {\bf k}} (a_{\bf k}^{\dagger} \sigma_{-,i}+ a_{\bf k}\sigma_{+,i}) \nonumber\\
&&+ \sum_{i} \Omega_{i}\sigma_{x,i}\cos(\omega_{li} t) \nonumber\\
&&+ \sum_{i,j}\frac{\delta_{ji}}{2} \sigma_{z,i}.
 \ea
The first term represents the individual energies $\omega_i$  of the QDs and $\sigma_{z, i}$ is the $z$-Pauli spin operator for QD $i$. The second term represents the photon energies $\theta_{\bf k}$ of the bath of photons with wave vectors ${\bf k}$; $a_{\bf k}^\dagger$ and $a_{\bf k}$ are photon creation and annihilation operators  The next term represents the energies $\xi_i$ of the classical charge traps with $\eta_i=+1$ for an occupied trap and $-1$ for an unoccupied trap. The next Jaynes-Cummings terms represent the interaction of the qubits with the bath of photons with coupling strength $\zeta_{i, {\bf k}}$. The next term describes the effect of two lasers, one coupled to each QD, where $\omega_{li}$ is the laser frequency and $\Omega_{i}$ is the coupling to QD $i$. The final term is the interaction $\delta_{ji}$ between the charge $j$ and QD $i$; $\delta_{ji}=1$ when the charge trap is occupied, and is zero otherwise.


We model our system by using a separate $4\times4$ density operator for each possible charge configuration, each of which evolve under the Hamiltonian parameters for that particular configuration. By stacking these on top of each other, we effectively create a rectangular matrix description of our system, as discussed in Appendix B: Thus altogether we have an object with $4\times (4 n)$ elements where $n$ is the number of possible charge trap states. When a charge fluctuates, we simply swap the density operators into a new order, reflecting the changed charge configuration. If the charges traps were quantum objects then we would require $(4n)\times(4n)$ elements to represent it - so we save a factor of $n$ in the state description by doing this.

%

In order to proceed we treat the bath of photons as weakly coupled to the QDs and express the dynamics of the QDs - fluctuators system in terms of a density matrix, leading to the  quantum optical master equation,\cite{breuer}
\begin{eqnarray}\label{master}
\frac{d}{dt} \rho(t) &=& -i[H_{I},\rho(t)] \nonumber\\
&&+ \sum_{i} {\gamma}_{i} \left(N_{i}+1 \right) \Gamma\big[{\sigma}_{-,i}, \rho(t) \big] \nonumber\\
&&+ \sum_{i} {\gamma}_{i} N_{i} \Gamma\big[{\sigma}_{+,i},\rho(t) \big]
\label{eq:ome}
\end{eqnarray}
where $H_I$ includes all but the Jaynes-Cummings and photon energy terms in Eq.~\ref{eq1}, following a transformation into a frame rotating with the two laser frequencies and a rotating wave approximation:
 \ba
 \label{eqRWA}
 H_I &=& \sum_{i}\frac{\nu_{i}}{2} \sigma_{z,i} +\sum_{j}\frac{\xi_{j}}{2}\eta_{j}\nonumber\\
&&+ \sum_{i} \frac{\Omega_{i}}{2}\sigma_{x,i} + \sum_{i,j} \frac{\delta_{ji}}{2} \mu_{ji},
 \ea
where $\nu_{i}\equiv\omega_i-\omega_{li}$ are the laser detunings. The dissipaters are given by
\be
\Gamma\big[ \hat{L}, \rho(t) \big] =  \left( \hat{L} \rho(t) \hat{L}^\dagger- \frac{1}{2} \hat{L}^\dagger \hat{L} \rho(t)  - \frac{1}{2} \rho(t) \hat{L}^\dagger \hat{L} \right).
\ee
The parameters in Eq.~\ref{master} are $\gamma_{i}$, which is the optical decay rate for the QD $i$ and $N_i$, which is the Bose-Einstein occupation number of the photon bath taken at the transition frequencies of QD $i$.

Our calculation proceeds using a transfer matrix approach. This is performed by first writing Eq.~\ref{master} in the form
\begin{equation}\label{eq23}
\dot{\rho}(t) = M \rho(t),
\end{equation}
where $M$ is the super-operator acting on the density operator $\rho$.
Since we are solving the system numerically and for small time steps, we can write
\begin{equation}\label{eq24}
\rho(t+\Delta t) = (1 + M\Delta t) \rho(t).
\end{equation}
This tells us that the transfer matrix for a system that does not fluctuate, but has a constant charge bias can be defined as $1 + M\Delta t$. We can now introduce the fluctuating nature of the charge by using a Pauli-$x$ operator, $\sigma_{xc}$ (where the `c' indicates that it operates on a charge) to flip the charge between its two states (see Appendix~\ref{appb}). We then divide the time steps into two parts: we assume that there is a probability $P \Delta t$ within the time $\Delta t$ that there is a change in the fluctuator's state, and a probability that there will be no change $1-P \Delta t$.\cite{joynt} In this paper, we will restrict the model to having the same rates for hopping onto and away from the trap sites correspondence to a temperature higher than the energy gap of the fluctuators. If $\Delta t$ is small then higher order terms can be neglected and we find that the  $(1 + M\Delta t)$ term in Eq.~\ref{eq24} is modified to:
\begin{equation}\label{transfer}
(1-P\Delta t)(1+M\Delta t) + P \Delta t \sigma_{xc} \approx 1+(M-P + P\sigma_{xc} )\Delta t.
\end{equation}
Eq.~\ref{transfer} defines the transfer matrix for fluctuating charges and with this we can calculate the dynamics of the system. 

\section{Calculating Intensity Correlations}\label{methods}

From the dynamical simulations of the system density matrix, we obtain predictions of emitted photon correlations. To this end we calculate the two-photon intensity correlation function
\begin{equation}\label{eq2}
g^{(2)} (t,t+ \tau ) = \frac{\langle  a_{1}^{\dagger}(t )a_{2}^{\dagger}(t+\tau )a_{2}(t+\tau) a_{1}(t)  \rangle}{\langle  a_{1}^{\dagger}(t) a_{1}(t) \rangle \langle a_{2}^{\dagger}(t+\tau )a_{2}(t+\tau)   \rangle}.
\end{equation}
The operators $a_i$ and $a_i^{\dagger}$ refer to the photon field detected by a detector $i\in{1,2}$ (see Fig.~\ref{stark}). Each detector is responsive only to a range of frequencies around the resonant frequency of each QD, and for QDs that are sufficiently detuned from one another we expect that photons which activate detector $i$ originate from QD $i$ only. The QDs in a typical experimental set up would be quite closely spaced and so a 
Hanbury Brown-Twiss set up \cite{brown} could be used. In this experiment, the photons from the QDs are passed through a beam splitter so that there are two paths that each lead to a detector, before which filters and polarisers can be placed depending on what exactly the experiment requires. The detection of a photon in the first detector begins a timer, which is then stopped by a detection of a photon in the second detector.\cite{fox} The results are collected into a histogram to display the number of events as a function of time between detection events.

The cross-correlation in Eq.~\ref{eq2} is in terms of photon creation and annihilation operators, but we can relate these field operators to our system operators through input-output theory.\cite{walls} In general an output field is a sum of contributions from an input field and from the decay of the systems (QDs in our case) that decay optically: $a_{out} (t)=a_{in} (t)+ \sqrt{\gamma_1} \sigma_{-,1} (t)+ \sqrt{\gamma_2} \sigma_{-,2} (t)$. We presume a typical setup in which $a_{in}$ is in the vacuum state, and so the output field is then the sum of two well frequency-resolved fields, which will be separately detected. We can then associated each detector field with a particular system operator:

\begin{equation}\label{eq13}
 a_i(t) = \sqrt{\gamma_i} \sigma_{-, i}(t).
\end{equation}
This leads to an expression for $g^{(2)}(t,t+ \tau ) $ which depends only on the QD system operators:\cite{carm}
\begin{equation}\label{eq14}
g^{(2)} (t,t+ \tau ) = \frac{\langle \sigma_{+,1}(t) \sigma_{+,2}(t+\tau)  \sigma_{-,2}(t+\tau ) \sigma_{-,1}(t) \rangle}{\langle \sigma_{+,1}(t) \sigma_{-,1}(t) \rangle \langle \sigma_{+,2}(t+\tau)  \sigma_{-,2}(t+\tau ) \rangle}.
\end{equation}

A general way of finding the two time correlation function from the master equation is to exploit the quantum regression theorem.\cite{scully} However, we can user a simpler method if we assume that experimental measurements take place when the system has reached a steady state ($ss$) with density operator $\rho_{ss}$. In this case $g^{(2)}$ is only dependent on the delay time $\tau$:
\ba
g^{(2)} (\tau)& =&\frac{\langle \sigma_{+,1}(0) \sigma_{+,2}(\tau)  \sigma_{-,2}(\tau ) \sigma_{-,1}(0) \rangle_{ss}}{\langle \sigma_{+,1} \sigma_{-,1} \rangle_{ss} \langle \sigma_{+,2}  \sigma_{-,2} \rangle_{ss}}\nonumber\\
&=& \frac{\tr[ \sigma_{+,1}(0) \sigma_{+,2}(\tau)  \sigma_{-,2}(\tau ) \sigma_{-,1}(0)  \rho_{ss}]}{\langle \sigma_{+,1} \sigma_{-,1} \rangle_{ss} \langle \sigma_{+,2}  \sigma_{-,2}\rangle_{ss}}.
\ea
 
Owing to the cyclicity of the trace we can project the steady state density matrix into the ground state:
\be\label{27}
g^{(2)} (\tau)=
\frac{\tr[\sigma_{+,2}(\tau)  \sigma_{-,2}(\tau )  (_1\bra{1} \rho_{ss} \ket{1}_1) \ket{0}_1{}_1\bra{0}]}{\langle \sigma_{+,1} \sigma_{-,1} \rangle_{ss} \langle \sigma_{+,2}  \sigma_{-,2}\rangle_{ss}},
\ee
where $\ket{1}_1$ and $\ket{0}_1$ represent the two basis state vectors for QD 1. We may now define a new density operator $\rho_P$, which is properly normalised and represents the steady state projected from the excited state of QD 1 to its ground state:
\be
\rho_P = \frac{_1\bra{1} \rho_{ss} \ket{1}_1\ket{0}_1{}_1\bra{0}}{\langle \sigma_{+,1} \sigma_{-,1} \rangle_{ss}} =  \ket{0}_1{}_1\bra{0}
\ee
and write
\be
g^{(2)} (\tau) =   \frac{\tr[\sigma_{+,2}(\tau)  \sigma_{-,2}(\tau ) \rho_P]}{{ \langle \sigma_{+,2}  \sigma_{-,2}\rangle_{ss}}} = \frac{\tr[\sigma_{-,2} \rho_P(\tau)\sigma_{+,2}  ]}{{ \langle \sigma_{+,2}  \sigma_{-,2}\rangle_{ss}}}.
\ee
After a long enough period, $\rho_P(\tau)$ becomes $\rho_{ss}$ and so $\lim_{\tau\rightarrow\infty}[g^{(2)} (\tau)] =1$ as expected.

\section{Results}\label{results}

In this section we present the $g^{(2)}(\tau) $ results across a variety of parameters and for a single fluctuator or two fluctuators.

\subsection{One charge fluctuator}

We begin with the case of a single fluctuator. Our aim is to assess what kinds of photon cross correlation signatures are obtained for different charge fluctuation rates and interaction strengths. As a starting point, we will assume there is a single charge fluctuator that affects each of the two qubits in the same way - i.e. $\delta_{11}=\delta_{12}$ in Eq.~\ref{eq1}.
We use QD parameters typical of InGaAs structures: Fixed throughout the paper will be the spontaneous decay rates ($\gamma_1=\gamma_2=1$ GHz), and the Rabi frequencies ($\Omega_1=\Omega_2 = 1$~GHz). Other parameters are varied for particular sets of results but their default values will be: Laser detuning $\nu_1=\nu_2=0$, charge fluctuation rate is $P = 1$~MHz. 
Using values of $0.8$ nm for the permanent dipole and $-34$~nm$^{2}$/V for the polarizability,\cite{starkrap} for our default choice of charge interaction strength $\mu_{11}=\mu_{12}=1$~GHz, we would need a charge at a distance of $1.32~\mu$m. As the charge is brought closer to the qubits it detunes the qubits either towards or away from the laser frequency depending on the initial size and direction of detuning. 

\begin{figure}[H]
\includegraphics[height=50mm]{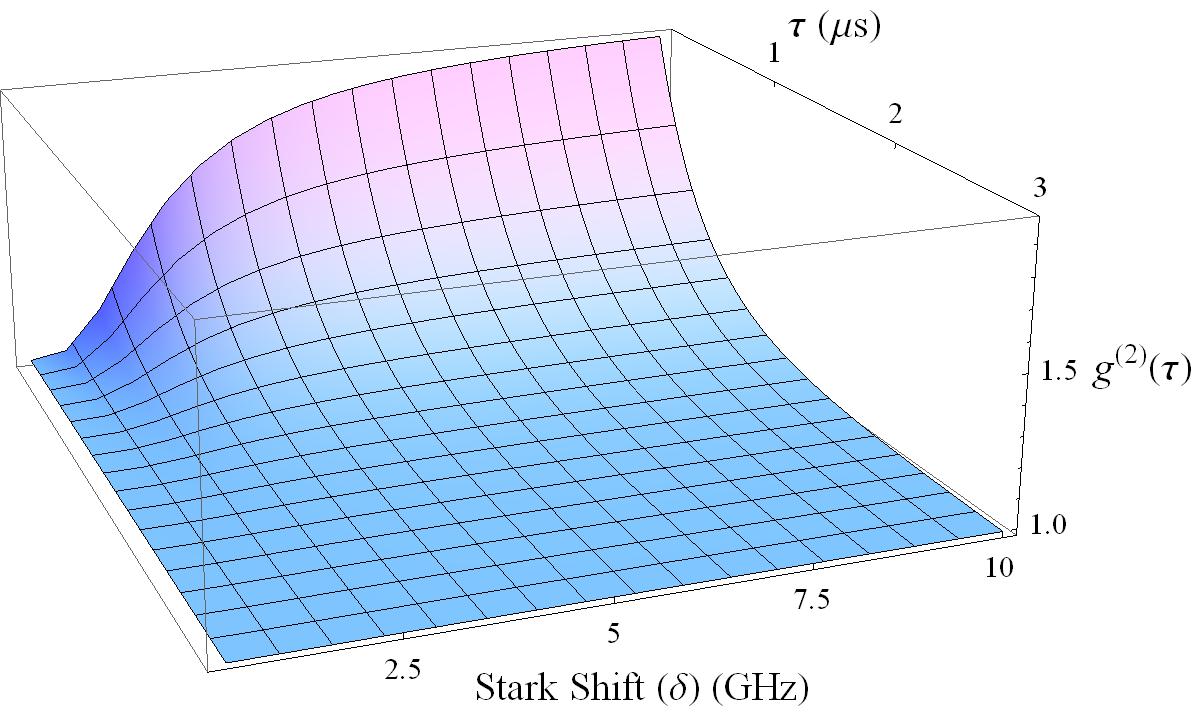}
\caption{$g^{(2)}(\tau)$ cross-correlation for two qubits interacting with a single charge, where the charge-qubit interaction strength ($\delta_{ij}$) is being varied. $\gamma_i$, the spontaneous decay rate of the qubits, is set to 1~GHz and determines the timescale of the return to uncorrelated photons.}\label{charge_strength}
\end{figure}

In Fig~\ref{charge_strength}, we show $g^{(2)}(\tau)$ as a function of $\mu_{11}=\mu_{12}$ between $0$ and $10$ GHz. This corresponds to effective detunings of $0$ and $6~\mu$eV, which encompasses the photoluminescence range from resonance to effectively zero photon emission for InGaAs type QDs.\cite{charge4}
We can see that the greater the charge-qubit interaction, the greater the initial cross-correlation of the detected photons. This initial correlation then decays back to the no-correlation value of $g^{(2)}=1$. For all values of the interaction strength this decay is on the 1~$\mu$s scale, which corresponds to the charge fluctuation rate. The explanation for this is straightforward: for a larger interaction, then either QD is only likely to emit when the charge trap is empty. If one QD emits, then since the noise is correlated, the other is likely to also emit -- at least for over a timescale less than the charge fluctuation time. On the other hand, for a smaller interaction -- less than the Rabi frequency $\Omega$ -- then it is also possible for a QD to be excited when the charge trap is occupied: an so we expect no cross correlations for $\mu_{1i}\ll\Omega_i$.

\begin{figure}[H]
\includegraphics[height=50mm]{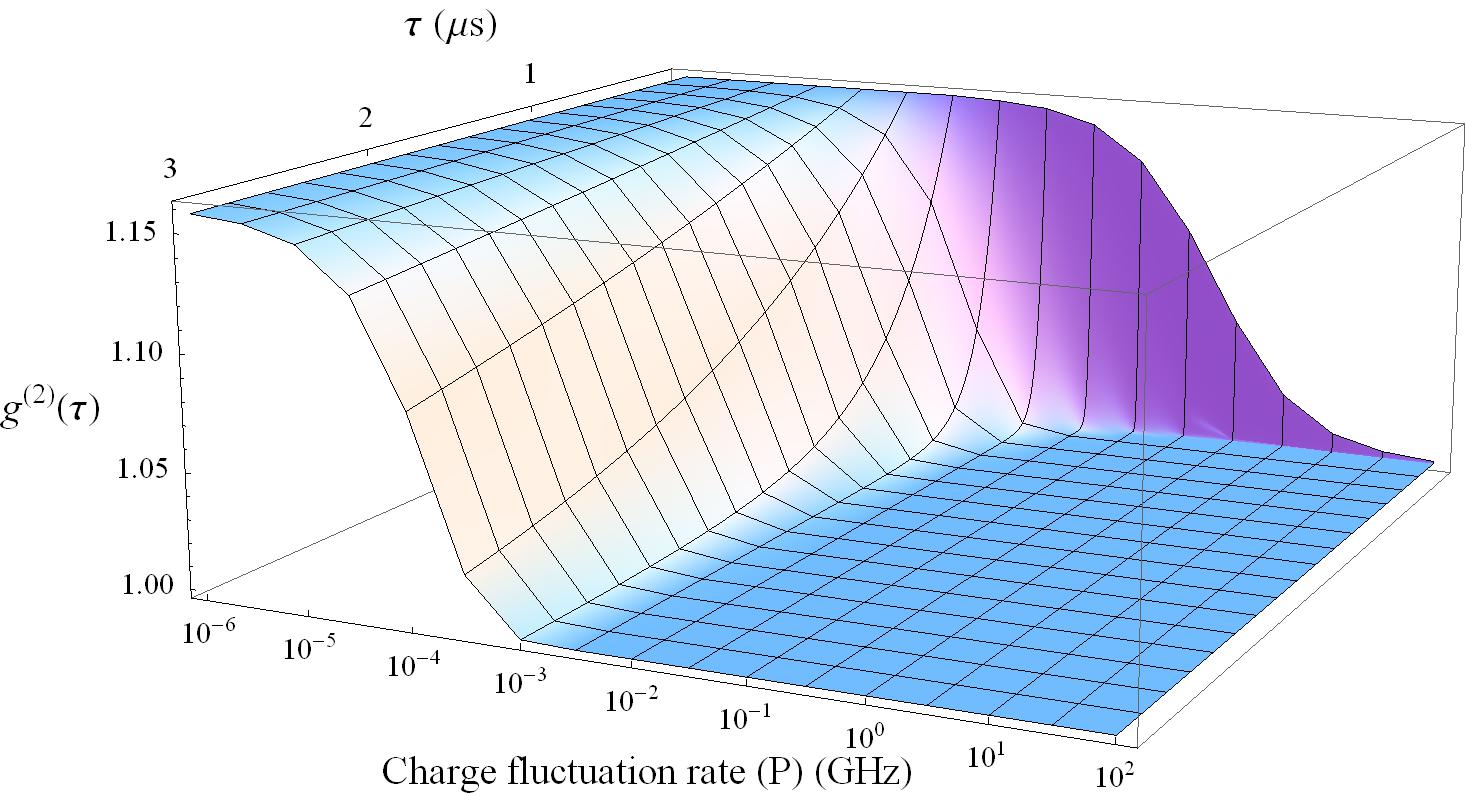}
\caption{$g^{(2)}(\tau)$ cross-correlation for two qubits interacting with a single charge, where the fluctuation rate ($P$) of the charge is being varied. The values of $P$ are given relative to $\gamma_i$ (the spontaneous decay rate of the qubits).}\label{charge_fluc}
\end{figure}

We next keep the charge-qubit interaction constant, and in Fig.~\ref{charge_fluc} look at how a changing fluctuation rate affects $g^{(2)}(\tau)$. When the fluctuation rate is smaller than the photon emission rate, we simply find that $g^{(2)}(\tau)$ decays on a timescale similar to that of the charge fluctuation rate. At the fastest fluctuation rates studied, however, there is a decrease in the initial value $g^{(2)}(0)$; this happens when the charge fluctuation rate exceeds the photon emission rate. In this regime, the experiment is no longer sensitive to the charge fluctuator.

\begin{figure}[H]
\includegraphics[height=45mm]{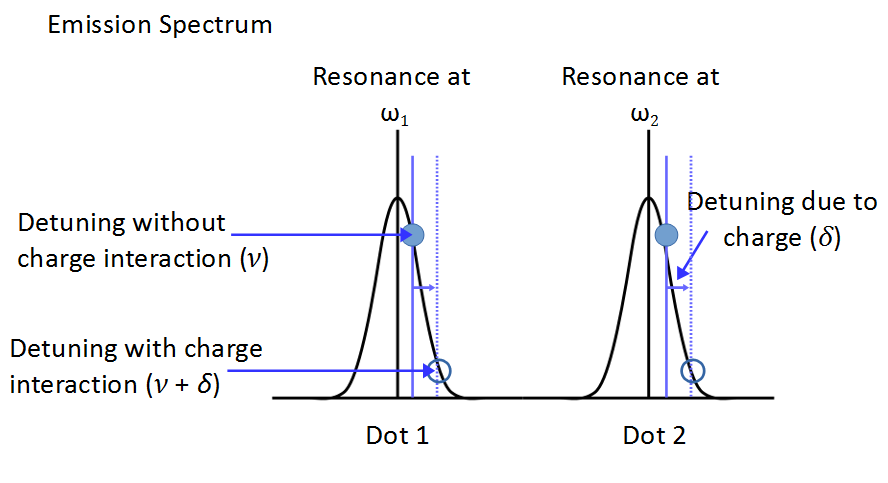}
\caption{Schematic drawing of the shifting detuning caused by charge interactions when the lasers are detuned in the same direction.}\label{det2}
\end{figure}

Since the effect of a charge fluctuator is to shift the resonance frequency of the qubits away from that of the lasers, we can use laser detuning as a further probe of the fluctuator correlation dynamics. 
In Fig.~\ref{det2} we illustrate a typical photoluminescence spectrum of two QDs. We know that if the two lasers are resonant with the two QDs then we expect to see correlated emitted photons  as shown in Fig.~\ref{charge_strength}. Imagine instead allowing both the lasers to be detuned from the QD resonance (see Fig.~\ref{det2}), by the same amount and in the same direction. Depending on the direction of the detuning, the effect of the charge will be to bring the QDs back into resonance or take them further from resonance. In this way, the emitted photons, regardless of the direction of detuning, will display a positive cross correlation.

\begin{figure}[H]
 \includegraphics[height=45mm]{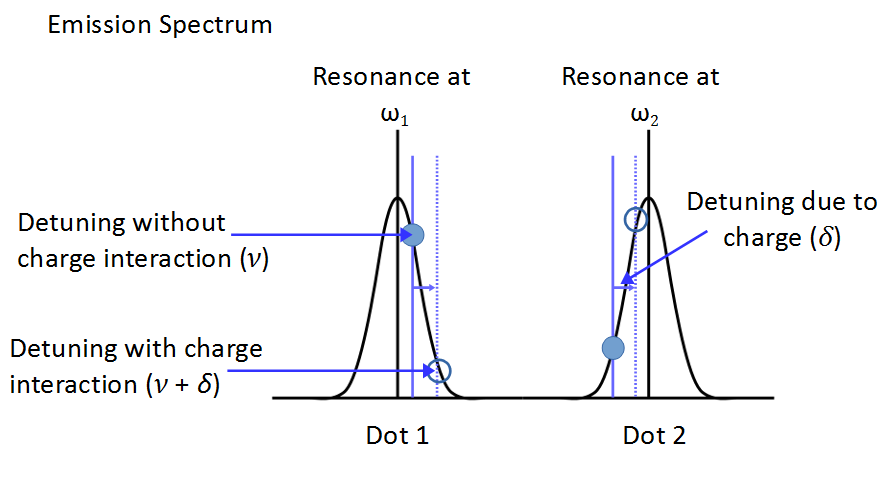}
 \caption{Schematic drawing of the shifting detuning caused by charge interactions when the lasers are detuned in opposite directions.}\label{det1}
 \end{figure}

On the other hand if the lasers are detuned in opposite directions, i.e. if one is blue shifted and the other red shifted with respect to the QD resonant frequencies, then charging the trap will shift one QD towards resonance and the other further away from it. This situation is shown in Fig~\ref{det1}. In this way the QD closer to resonance is more likely to emit a photon, while the other is less likely. We then expect a negative correlation between the emitted photons, for a large enough initial detuning.

\begin{figure}[H]
\includegraphics[height=50mm]{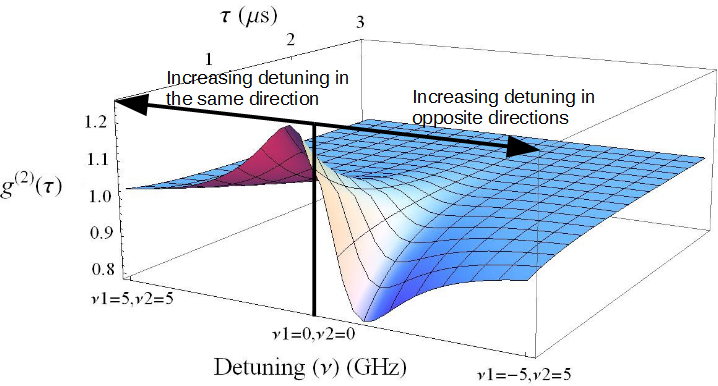}
\caption{$g^{(2)}(\tau)$ cross-correlation for two qubits interacting with a single charge, where the laser detuning ($\nu_i$) is being varied. On the left of the zero detuning line, detuning increases and is equal in magnitude and sign for both dots; on the right it increases with equal magnitude but opposite signs. }\label{1charge_det}
\end{figure}

In Fig~\ref{1charge_det} we show the cross-correlation for both the detuning scenarios just described. As expected, we see positive correlation for same sense detuned, and negative for opposite sense -- expect around $\nu=0$, when the charge fluctuation moves both QDs away from resonance and we recover a positive cross-correlation.

\subsection{More than one charge}

Let us now introduce a second charge fluctuator into the model, and establish whether it is possible to distinguish multiple from single fluctuators by using cross correlation measurements.

\begin{figure}[H]
\includegraphics[height=50mm]{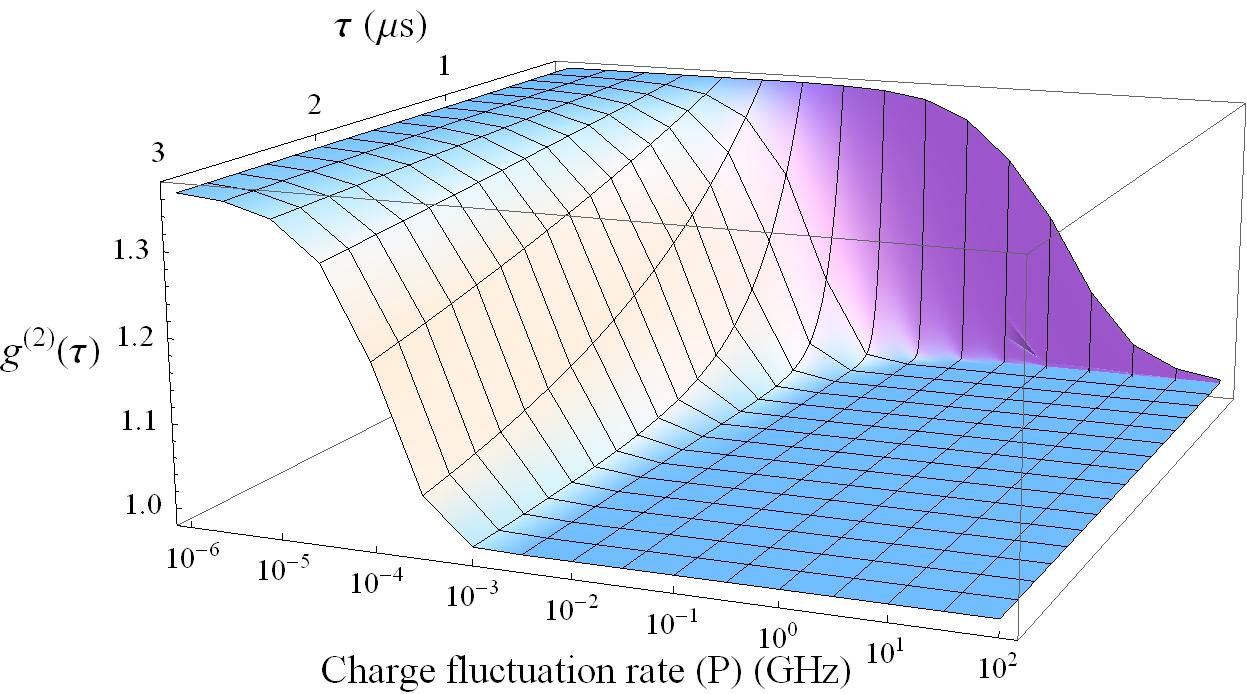}
\caption{$g^{(2)}(\tau)$ cross-correlation for two qubits interacting with two charges, where the fluctuation rate ($P_i$) of the charges are being varied. The values of $P$ are set equal to each other.}\label{2charge_fluc}
\end{figure}

\begin{figure}[H]
\includegraphics[height=50mm]{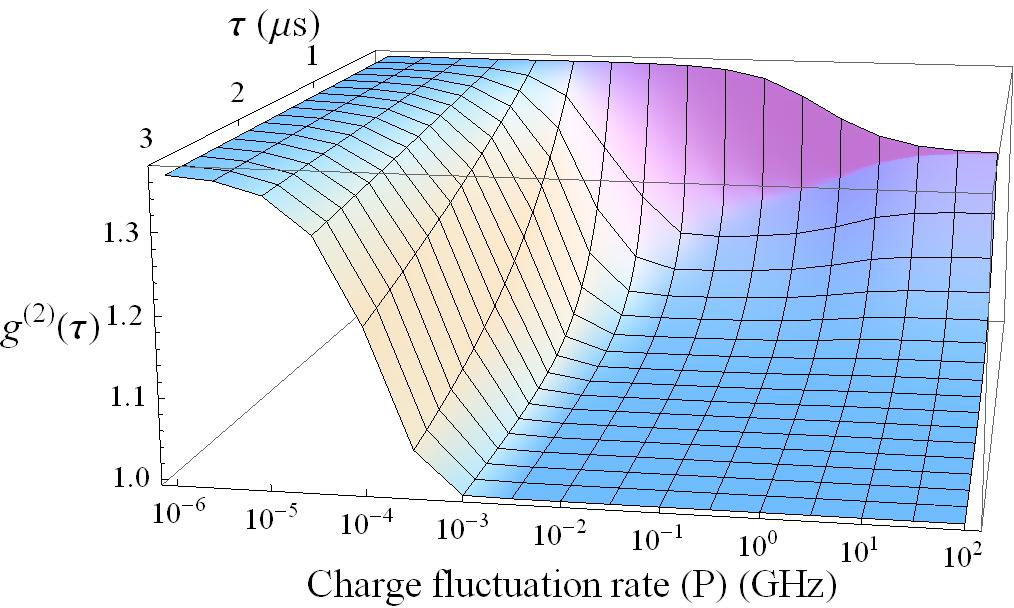}
\caption{$g^{(2)}(\tau)$ cross-correlation for two qubits interacting with two charges, where the fluctuation rate ($P_i$) of one charge is being varied and the other held at $10^{-3}$ GHz. }\label{2charge_fluc2}
\end{figure}

Fig.~\ref{2charge_fluc} shows $g^{(2)}(\tau)$ as a function of (equal) fluctuation rates for the two charges. We can see immediately that $g^{(2)}(0)$ is larger in this case than for the single fluctuator case shown in Fig.~\ref{charge_fluc}. This is expected since the two fluctuators working together increase the total possible detuning of each QD; as we have seen in Fig.~\ref{charge_strength} this results in a higher initial cross-correlation. As the fluctuation rate increases, the decay of $g^{(2)}(\tau)$ happens at shorter times, similar to the single charge case. 

\begin{figure}[H]
\includegraphics[height=100mm]{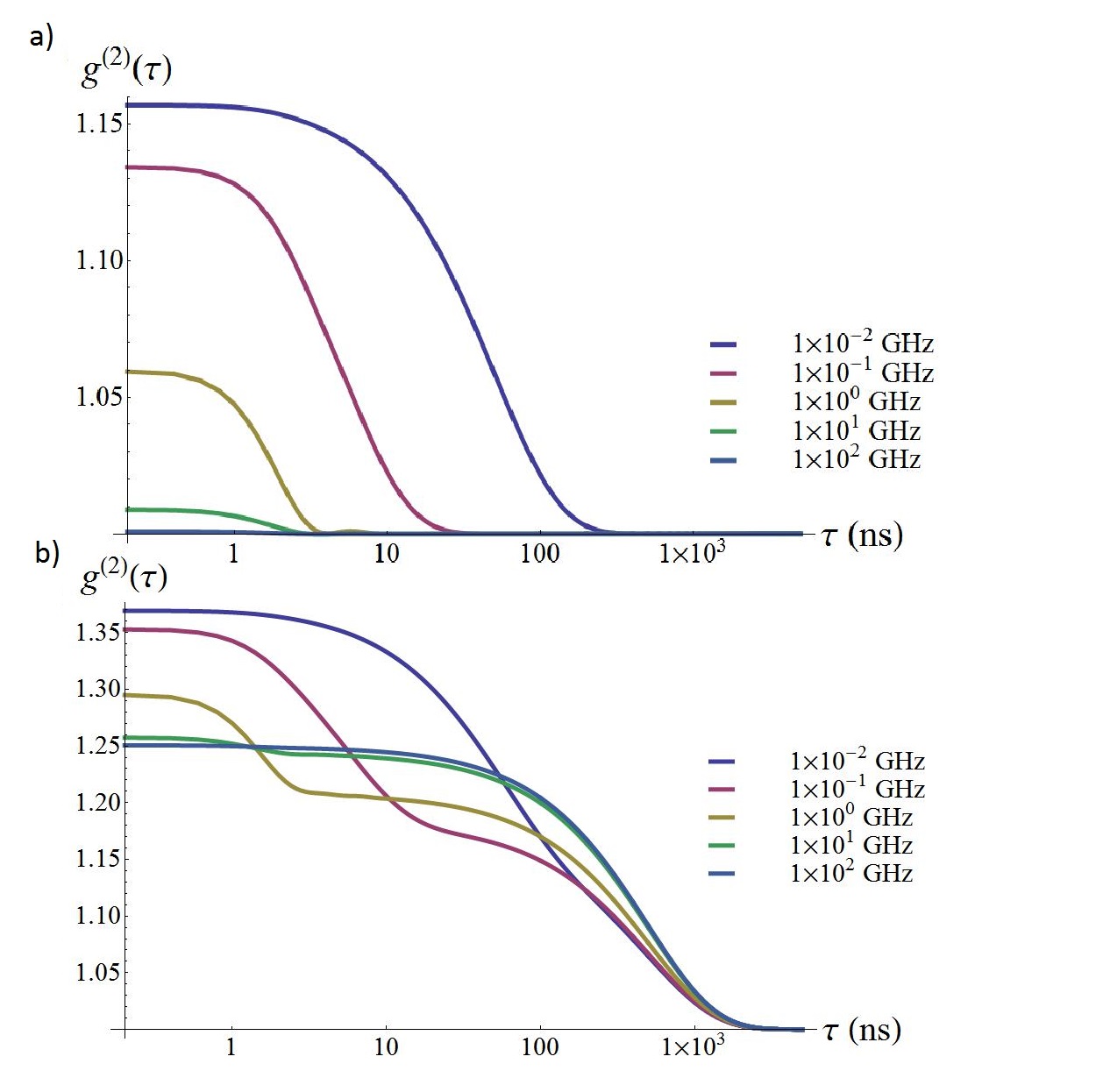}
\caption{Log time plots for some representative frequencies of the data shown in Fig.~\ref{charge_fluc} shown in a) and Fig.~\ref{2charge_fluc2} shown in b).}\label{log_1c}
\end{figure}

It is unlikely, however, that the two charge traps will be fluctuating at exactly the same rate, so let us now look into how different rates of fluctuation affect the cross-correlation. In Fig.~\ref{2charge_fluc2}, we show the impact on $g^{(2)}(\tau)$ of altering one fluctuation rate while keeping the other fixed at 10$^{-3}$~GHz. Comparing this figure with that for a single fluctuator shown in Fig.~\ref{charge_fluc}, we find that in certain cases it is possible to see a clear qualitative difference between the results for a single and two fluctuators. This is easier to discern by taking cuts through the plots for particular fluctuation rates, and using a log scale for the time; for the single and two fluctuator cases these are shown in Fig.~\ref{log_1c}(a) and Fig.~\ref{log_1c}(b). If the rates for the two fluctuators are significantly different then two plateaux can be seen in the curves, with two decay rates corresponding to two different fluctuation rates; this effect washes out once the faster fluctuation rate approaches that of the QD optical decay rate.

As would be expected, when the fluctuation rates of the charges are very similar, no deviation from the single fluctuation curves can be distinguished: They do not exhibit the plateaued structure that can be seen in Fig~\ref{log_1c} b). 

\begin{figure}[h]
\includegraphics[height=60mm]{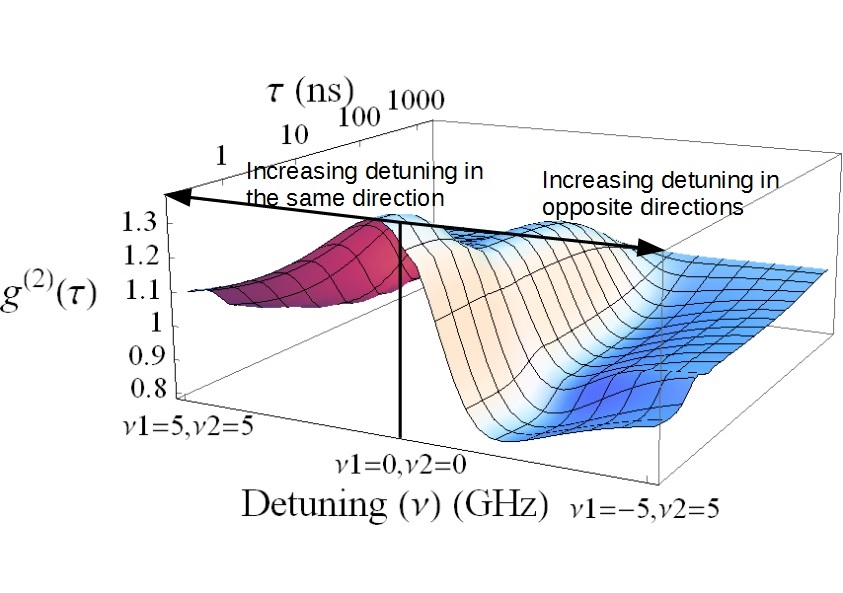}
\caption{$g^{(2)}(\tau)$ cross-correlation for two qubits interacting with two charges, where the laser detuning ($\nu_i$) is being varied. The two charges fluctuate with different rates of 1~MHz and 1~GHz.}\label{2charge_det}
\end{figure}

\begin{figure}[h]
\includegraphics[height=45mm]{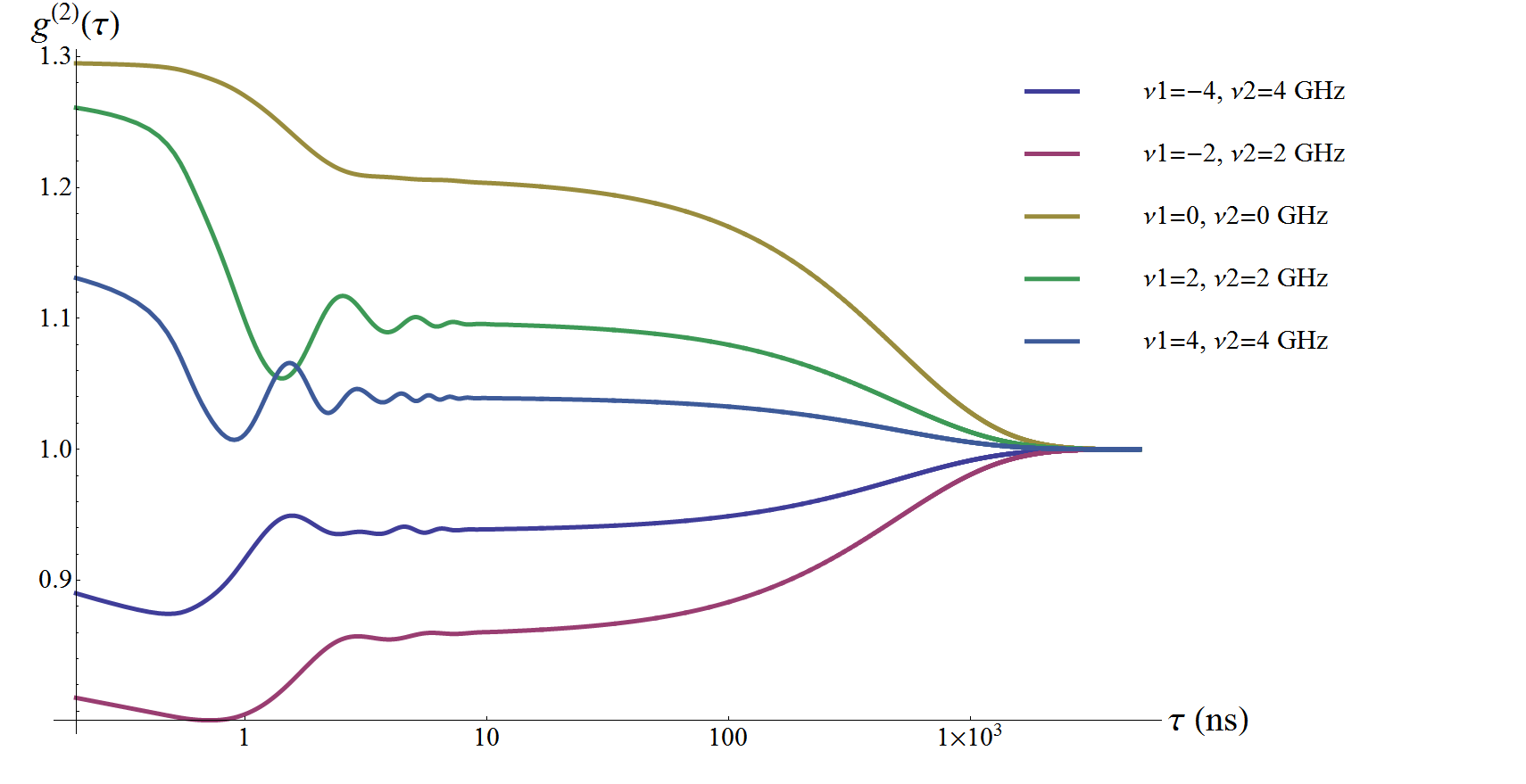}
\caption{Cuts through Fig.~\ref{2charge_det}, for five values of the detuning parameters.}\label{fig:cuts2charge}
\end{figure}

Finally, we look at varying the laser detunings for two qubits and two charges. In Fig~\ref{2charge_det} we show the cross correlation function for the same detuning parameters as in Fig.~\ref{1charge_det}, in the case where the two fluctuation rates are not equal (1~GHz and 1~MHz). We also display various cuts through this 3D plot, for different values of the detuning, in Fig.~\ref{fig:cuts2charge}. There is a clear contrast here with the surface shown in Fig.~\ref{1charge_det}; the cuts in Fig.~\ref{fig:cuts2charge} show that the plateaued structure characteristic of two different rates survives as detuning is varied. At negative detuning the a negative correlation is observed, but with a clear long time plateau. The oscillatory behaviour as detuning increases is a consequence of the increasing effective Rabi frequency, which is no longer fully damped by the
1~GHz optical decay processes.

\section{Summary}\label{summary}

We have described a measurement which can reveal information about the long range correlations in the decohering environment of a controlled quantum system. We have shown that for an environment consisting of one or two fluctuating charge traps that the time dependence and initial value of the cross-correlation generate a signature of a correlated environment and in some cases can give an indication of how many charge traps are present. 

Figs.~\ref{1charge_det} and \ref{2charge_det} display the variation of the experimental signature as a function of the detuning. Since detuning can be varied in a single experiment, these are key predictions that provide the most experimentally accessible signature of correlation, and clear differences are observed for the single and two charge trap cases. The switching from a positive to a negative correlation as a function of detuning serves as a distinctive signature of a common environment; the double plateaued structure is a sign of two distinct fluctuation rates.

As the numbers of charges increase beyond two, several changes are expected to the $g^{(2)}(\tau)$ plots. The initial correlation of the emitted photons will be higher, due to the larger number of charges giving an effective greater charge interaction strength in line with Fig. \ref{charge_strength}. Additionally, if the charges have widely varying fluctuation rates, then each charge would be seen as a separate plateau. 

Experiments which detect the effects we have described would not rely on any fast detectors or particularly fast optics. They are relatively simple measurements that could be performed on a suitable sample immediately, and would provide a unique probe of the unexplored collective effects of open system environments.

\acknowledgements

VP is supported by the EPSRC Scottish Doctoral Training Centre in condensed matter physics (EP/G03673X/1). BWL thanks the Royal Society for a University Research Fellowship.

\appendix
\section{Qubits and Charge Hamiltonian}
\label{appa}
\noindent
Here we present the Hamiltonian construction in the case of two QDs and a single charge; more complex situations follow similarly. Each QD $i$ ($i\in \{1, 2\}$) has a ground state $\ket{0}_i$ and excited state as $\ket{1}_i$, so that their uncoupled Hamiltonian is denoted in a spin language as $\sigma_{z, i}=\ket{1}_i\bra{1} - \ket{0}_i\bra{0}$. The charge fluctuator two level systems are classical and so cannot have coherences. While a quantum two level system is represented as a two by two density matrix, a classical object has no coherences, so the off-diagonal matrix elements are unnecessary. A two-vector of classical populations ($\eta_{j}$) then describes the state of fluctuator $j$, with the first element giving the population in the ground (no-charge) state, and the second that in the excited (charged) state.
If we use the usual tensor product formulation of quantum mechanics of more than one subsystem, our overall description of our qubit-fluctuator state must then be a rectangular matrix, corresponding to 2$^N$ stacked square matrices, one for each of the classical states of $N$ fluctuators. Between charge fluctuation events each square density operator acts independently of the others and so each can be treated individually. For example, the case of two qubits and a single charge, which is described by a $4\times8$matrix, can be treated as two $4\times4$ matrices, each corresponding to one of the states of the charge. 

With this in mind we define the system Hamiltonian as

\begin{equation}
H_{S} = \frac{1}{2}\omega_{1} \sigma_{z,1} + \frac{1}{2}\omega_{2}\sigma_{z,2} + \frac{1}{2}\xi_{1}\eta_{1},
\end{equation}
where we use
\begin{eqnarray}\label{vecs}
\sigma_{z,1}&=&\left(\begin{array}{c}
1\\
1
\end{array}\right) \otimes \mathbb{1}_{2} \otimes \sigma_{z},\nonumber\\
\sigma_{z,2}&=&\left(\begin{array}{c}
1\\
1
\end{array}\right) \otimes \sigma_{z} \otimes \mathbb{1}_{2},\nonumber\\
\eta_{1}&=&\left(\begin{array}{c}
1\\
-1
\end{array}\right) \otimes \mathbb{1}_{2} \otimes \mathbb{1}_{2}.
\end{eqnarray}
The first column vector describes the charge state; $\omega_{i}$ is energy required to excite uncoupled QD $i$ from ground to excited state. $\xi$ is the energy difference between the charge being at the trap site and the charge being elsewhere, which we assume to be at an infinite distance. Each operator is now of dimension $4\times8$; during coherent evolutions the top and bottom square matrices of this object are treated individually and act on the top and bottom square matrices of the corresponding density operator respectively, following the usual rules of quantum mechanics. Operator eigenstates corresponding to the different charge configurations can be found for each square component matrix.

\section{Interaction Hamiltonian and Fluctuations}
\label{appb}

The qubit-bath interaction Hamiltonian is defined as:
\begin{equation}
H_{q,B} = \sum_{\bf k}\zeta_{1,{\bf k}} (a_{\bf k}^{\dagger} \sigma_{-,1}+ a_{\bf k}\sigma_{+,1}) +  \zeta_{2,{\bf k}}
(a_{\bf k}^{\dagger} \sigma_{-,2}+ a_{\bf k}\sigma_{+,2}),
\end{equation}
where
\begin{eqnarray}
\sigma_{+,1}&=&\left(\begin{array}{c}
1\\
1
\end{array}\right) \otimes \mathbb{1}_{2} \otimes \sigma_{+}, \nonumber\\
\sigma_{+,2}&=&\left(\begin{array}{c}
1\\
1
\end{array}\right) \otimes \sigma_{+} \otimes \mathbb{1}_{2},   \nonumber\\
\sigma_{-,1}&=&\left(\begin{array}{c}
1\\
1
\end{array}\right) \otimes \mathbb{1}_{2} \otimes \sigma_{-},  \nonumber\\
\sigma_{-,2}&=&\left(\begin{array}{c}
1\\
1
\end{array}\right) \otimes \sigma_{+} \otimes \mathbb{1}_{2}.  
\end{eqnarray}
The interaction here describes the couplings between the QDs and the photonic bath, with coupling strengths $\zeta_{1, {\bf k}}$ or $\zeta_{2, {\bf k}}$, where $a_{\bf k}$ and $a_{\bf k}^{\dagger}$ are the annihilation and creation operators for the photon bath mode with wave vector ${\bf k}$. The QD raising and lowering operators in the interaction Hamiltonian are denoted by the $\sigma_{+}$ and $\sigma_{-}$ Pauli matrices, which are constructed as $\ket{1}\bra{0}$ and $\ket{0}\bra{1}$ respectively. The effect of this part of the interaction with the system is taken into account by deriving a Born-Markov optical master equation.

We also account for a coupling between the qubits and the (classical) lasers. The qubit-laser Hamiltonian is
\begin{equation}
H_{q,l} = \Omega_{1}\sigma_{x,1}\cos(\omega_{l1} t)+\Omega_{2}\sigma_{x,2}\cos(\omega_{l2} t),
\end{equation}
where $\Omega_i$ is the Rabi frequency of the $i$th QD and $\omega_{li}$ is the laser frequency of laser $i$ which is assumed to drive only the $i$th QD with which it is closely resonant. Similarly to previous definitions:
\begin{eqnarray}
\sigma_{x,1}&=&\left(\begin{array}{c}
1\\
1
\end{array}\right) \otimes \mathbb{1}_{2} \otimes \sigma_{x}, \nonumber\\
\sigma_{x,2}&=&\left(\begin{array}{c}
1\\
1
\end{array}\right) \otimes \sigma_{x} \otimes \mathbb{1}_{2} .
\end{eqnarray}

The final interaction is between the qubits and charge:
\begin{equation}
H_{q,c} = \frac{\delta_{11}}{2}\mu_{11}+\frac{\delta_{12}}{2} \mu_{12},
\end{equation}
where
\begin{eqnarray}
\mu_{11}&=&\left(\begin{array}{c}
1\\
0
\end{array}\right) \otimes \mathbb{1}_{2} \otimes \sigma_{z}, \nonumber\\
\mu_{12}&=&\left(\begin{array}{c}
1\\
0
\end{array}\right) \otimes \sigma_{z} \otimes \mathbb{1}_{2}.
\end{eqnarray}
There are two terms, which show the interaction of the charge (labelled 1 here since we are considering only a single charge) with both of the QDs. In general the interaction is $\delta_{ji}$, where the subscript denotes an interaction of charge $j$ with QD $i$. The term describes a process in which the stationary charge at some well defined distance introduces a Coulomb potential that gives rise to a Stark shift of the excitonic states, thus creating a TLS with a larger energy spacing.

The qubit-laser Hamiltonian is time dependent, but we can remove that time dependence by moving into a rotating frame and performing the rotating wave approximation (RWA). This is done by applying a unitary transformation to all part of the Hamiltonian that involve the dots and the charges, i.e., all parts except $H_{q,B}$. From the time dependent Schr\"odinger equation, we get:
\begin{eqnarray}
-i \hbar \partial_{t} (U^{-1} \mid \psi \rangle ) &=& -i \hbar (\partial_{t} U^{-1})\mid \psi \rangle + U^{-1}H \mid \psi \rangle \nonumber\\
&=& H U^{-1} \mid \psi \rangle \nonumber
\end{eqnarray}
Therefore, the effective Hamiltonian becomes, after moving to the rotating frame
\begin{equation}
H_{\rm rf} = U H U^{-1} +U(i \hbar \partial_{t} U^{-1}).
\end{equation}

The operator $U^{-1}$ is 
\begin{equation}
\left(
\begin{array}{cccc}
 e^{\frac{1}{2} i t ({\omega_{l1}}+{\omega_{l2}})} & 0 & 0 & 0 \\
 0 & e^{\frac{1}{2} i t ({\omega_{l2}}-{\omega_{l1}})} & 0 & 0 \\
 0 & 0 & e^{\frac{1}{2} i t ({\omega_{l1}}-{\omega_{l2}})} & 0 \\
 0 & 0 & 0 & e^{\frac{1}{2} i t (-{\omega_{l1}}-{\omega_{l2}})} \\
\end{array}
\right).
\end{equation}

Notice that this is a $4 \times 4$ matrix. As stated previously, we can consider the two charge states as independent and as such, the $4 \times 8$ rectangular matrix can be thought of as two $4 \times 4$ matrices. This unitary transformation operates on each of those two parts individually. After performing the rotating wave approximation, which removes rapidly oscillating terms, i.e. those with twice the frequency of $\omega_{l1}$ and $\omega_{l2}$, and letting $\omega_{i} - \omega_{li} = \nu_{i}$, we get:
\begin{equation}
\begin{split}
H_{S}'=\frac{1}{2}\left(\nu_{1} \sigma_{z,1} + \nu_{2} \sigma_{z,2} + \Omega_{1} \sigma_{x,1} + \Omega_{2} \sigma_{x,2}\right)\\
+\frac{1}{2}\left(\frac{\delta_{11}}{2}\mu_{12}+\frac{\delta_{12}}{2} \mu_{22}+\xi_{1}\eta_{1}\right).
\end{split}
\end{equation}
This is now our final system Hamiltonian; in the paper we show how this is taken together with $H_{q, B}$ to find a quantum optical master equation for the qubit-charge system.

Additionally, we need to define an operator in which a single charge fluctuates from occupied to unoccupied or {\it vice versa}, an operation executed by the operator $\sigma_{xc}$. In our rectangular density operator notation for a single charge this simply corresponds to swapping over the upper and lower square matrices. For more complex situations of more charge fluctuators, a similar reordering of the now multiple square matrices achieves the desired effect.

\end{document}